%% file: main.tex
\documentclass[12pt]{article}
\input{aux/macros.tex}
\usepackage[affil-it]{authblk}

\title{A single-snapshot inverse solver for two-species graph model of tau pathology spreading in human Alzheimer's disease}

\author[1]{Zheyu Wen}
\author[1]{Ali Ghafouri}
\author[1]{George Biros\thanks{Corresponding author: biros@oden.utexas.edu}}

\affil[1]{Oden Institute, The University of Texas at Austin, Austin, Texas, USA}

\date{}

\begin{document}
\maketitle

\input abstract


\input intro

\input method

\input results

\input conclusion

\input acknowledgement_data_availability

 \bibliographystyle{vancouver} 
 \bibliography{refs}

\end{document}

%% file: aux/macros.tex
\usepackage{amsmath,amssymb,amsfonts}
\usepackage{graphicx}
\usepackage{textcomp}

\usepackage{enumitem}
\usepackage{tabularx,booktabs}
\newcolumntype{Y}{>{\centering\arraybackslash}X}
\usepackage{url}            
\usepackage{booktabs}       
\usepackage{nicefrac}       
\usepackage{microtype}      
\usepackage{lipsum}
\usepackage{bbm}
\usepackage{breqn}
\usepackage{pdfpages}
\usepackage{mathtools}
\usepackage{subcaption}
\usepackage{algorithm}
\usepackage[noend]{algpseudocode}
\usepackage{diagbox}
\usepackage{adjustbox}
\usepackage{float}
\usepackage{lipsum}                     
\usepackage{xargs}                      
\usepackage{bm} 
\usepackage{mathtools}
\usepackage{xcolor,colortbl}
\usepackage{lscape}
\usepackage{siunitx}
\usepackage{enumitem}
\usepackage{dsfont}
\usepackage[textsize=tiny,colorinlistoftodos]{todonotes}
\usepackage{hyperref}
\usepackage{cleveref}
\usepackage{dsfont}

\usepackage[T1]{fontenc}
\usepackage{lmodern}
\definecolor{Gray}{gray}{0.9}
\makeatletter
\newcommand{\crefnames}[3]{%
  \@for\next:=#1\do{%
    \expandafter\crefname\expandafter{\next}{#2}{#3}%
  }%
}
\makeatother
\crefnames{part,chapter,section}{\S}{\S\S}

%% file: abstract.tex
\begin{abstract}
We propose a method that uses a two-species ordinary differential equation (ODE)  model to characterize misfolded tau protein spreading in Alzheimer's disease (AD) and calibrates it from clinical data. 
The unknown model parameters are the initial condition (IC) for tau and three scalar parameters representing the migration, proliferation, and clearance of tau proteins. 
Driven by imaging data, these parameters are estimated by formulating a constrained optimization problem with a sparsity regularization for the IC. 
This optimization problem is solved with a projection-based quasi-Newton algorithm. 
We evaluate the performance of our method on both synthetic and clinical data. 
Patients are from the AD Neuroimaging Initiative (ADNI) datasets: 455 cognitively normal (CN), 212 mild cognitive impairment (MCI), and 45 AD subjects. 
We compare the performance of our approach to the commonly used Fisher-Kolmogorov (FK) model with a fixed IC at the entorhinal cortex (EC). 
Our method demonstrates an average improvement of 25.7\% relative error compared to the FK model on the AD dataset. 
HFK also achieves an R-squared score of 0.664 for fitting AD data compared with 0.55 from FK model results under the same optimization scheme. 
Furthermore, for cases that have longitudinal data, we estimate a subject-specific AD onset time.
\end{abstract}

%% file: intro.tex
\section{Introduction}
\label{sec:intro}
Misfolded tau is a neurotoxic protein that correlates with the progression of Alzheimer's disease (AD). It disrupts normal nervous system function and leads to brain atrophy \cite{frontzkowski2022earlier, blinkouskaya2021brain, franzmeier2020functional, yu2021human}. Emerging evidence suggests that abnormal tau propagates along neuronal pathways \cite{cope2018tau, jacobs2018structural}.

Understanding tau propagation and its impact on cognitive function remains a challenge. Positron Emission Tomography (PET) scans using the F-AV-1451 tracer (tauvid) visualize of tau protein spread \cite{johnson2016tau}. Several research groups \cite{vogel2020spread, fornari2019prion, scheufele2020calibration} have developed methods to process tau-PET images and establish temporal dynamic models that describe tau spreading in human brains. However, several challenges need to be addressed. First, tau-PET images are often contaminated by off-target signals. Second, the time gap between consecutive PET scans is relatively short compared to the entire progression timeline of AD. Third, PET scans do not provide conclusive information on where the initial misfolding takes place. 

Mathematical modeling can be used to analyze scans and test different hypotheses of tau evolution. The FK model~\cite{fisher1937wave, kolmogorov1937study} has been widely employed to describe tau dynamics, assuming that the initial seeding location is the entorhinal cortex (EC) \cite{vogel2020spread, fornari2019prion}. However, this model exhibits discrepancies with PET data, as it predicts high abnormal activity in the IC during the FK forward evolution. This contradicts observed PET scans, where the EC does not consistently exhibit significant activity. (For specific examples, refer to \Cref{sec:results} and \cite{devos2018synaptic, hoenig2018networks, vogel2021four}).

\subsection{Summary of the method}
To better understand the tau misfolding initiation location we propose a two species model. By fitting this model to longitudinal PET data, we can gain a better understanding of the progression of AD over time. To quantify the tau abnormality within each region of interest (ROI), we utilize the maximum mean discrepancy (MMD) metric~\cite{gretton2012kernel}. To infer the model's parameters, particularly the sparse IC of the temporal model, we present a robust inversion scheme tailored to fitting the model using a single snapshot data. 

\subsection{Contributions}
Our contributions can be summarized as follows:
\begin{enumerate}
\item We employ a two-species heterodimer Fisher-Kolmogorov (HFK) model, which considers both normal and abnormal tau \cite{fornari2019prion, schafer2019interplay}.
 We hypothesize that the abnormal tau originates in a small number of regions and we introduce an IC sparsity constraint. We propose an iterative inversion algorithm \cite{subramanian2022ensemble}. 
  \item We propose a scheme for extracting tau abnormality statistics from tau-PET scan MMD to quantify differences of standard uptake value ratio (SUVR) distributions in each brain region and cerebellum. We demonstrate that our method performs better than other metrics such as  regional mean SUVR and tau positive probability~\cite{vogel2020spread}.
  \item We conduct several experiments to assess the proposed methodology. We use synthetic data to verify the model. We also analyze algorithm sensitivity to parcellation, brain templates, and tractography methods. We apply the method to clinical data from the ADNI dataset~\cite{petersen2010alzheimer} and study
the inconsistencies in the FK model, the effect of inverting for the tau IC, and compare with other methods.
\end{enumerate}

\subsection{Related Work}
We first discuss methods for  quantifying abnormalities in tau-PET scans and then we discuss calibration of biophysical models use tau-PET scans.

\textit{Tau-abnormality from tau-PET}: In \cite{vemuri2017tau} the authors use the standard uptake value ratio (SUVR) by normalizing voxel tau PET images using the median tau level of the cerebellum. Subsequently, the mean SUVR signal is calculated for each region to represent the degree of tau abnormality within that region. It is well known, however, that tau PET images comprise not only abnormal tau signals but also off-target signals. To account for this, Vogel et al. \cite{vogel2020spread} propose a novel approach utilizing a two-component Gaussian Mixture Model (GMM) to filter out the off-target signal. This involves collecting tau voxel data from each ROI across all subjects in the dataset and fitting both one and two-component GMMs. If the two-component GMM provides a superior fit, the second component is leveraged to determine tau abnormality or tau probability through the Gaussian cumulative distribution function.  This method quantifies abnormalities based on the distribution across the entire population; as such it yields different results for the same subject, depending on the dataset.

\textit{Tau propagation modeling}: There are several studies on biophysical models for abnormal tau~\cite{raj2021graph}. Here we focus on models that have been calibrated using tau-PET data. The single-species FK model~\cite{vogel2020spread, scheufele2020calibration, cohen1994structural, jarrett1993seeding, bertsch2017alzheimer, fornari2019prion, weickenmeier2019physics} is one of the most popular ones. The model requires two scalar coefficients and the initial tau seeding locations to determine tau dynamics.  Its main shortcoming, as discussed in~\cite{wen2023two}, the FK model preserves the location of the tau maximum among highly abnormal regions in FK forward solution. Vogel \cite{vogel2020spread} uses the Epidemic Spreading Model, which is a FK variant with an additional clearance term. Regarding inversion of model parameters, Schafer \cite{schafer2021bayesian} takes bayesian and longitudinal tau-PET scan data to infer the parameter distribution. Vogal \cite{vogel2020spread} calibrates the model parameters by grid-search optimization with the EC region as the seeding region. Despite the development of a two-species model in \cite{fornari2019prion} and more complex models developed in \cite{thompson2020protein,pal2022nonlocal}, an inversion algorithm accounting both sparse initial condition and model operating parameters is needed.

The works in \cite{iturria2014epidemic} also investigated the effect of ICs using either an  exhaustive search with a pre-determined number of seeds or assuming that the IC was at regions with significant tau.  The authors \cite{vogel2021four} tried all left-right pairs cortical ROIs to be the epicenter and finally chose the best performed IC. This algorithm only chooses two regions as candidates of IC.

In this study, we focus on model calibration algorithms and use IC inversion algorithms adapted from \cite{needell2009cosamp,subramanian2020did,scheufele2020fully,subramanian2022ensemble}.

%% file: method.tex
\section{Methodology}
\label{sec: method}

Following~\cite{schafer2021bayesian,vogel2020spread,kim2019comparison}, we use a graph-based approach that coarsens the brain domain using standard parcellations and tractography-based connectivity. Let $\mathcal{G}$ be a graph whose vertices correspond to brain parcels/regions defined in a standard atlas; and whose edges are defined by an effective connectivity between these nodes. We define $\mathbf{c}_a(t)\in\mathbb{R}^N$ as a degree of tau \emph{abnormality} on the graph nodes at time $t$, with $c_a^i(t)\in \left[0, 1\right]$ at vertex $i$ in $\mathcal{G}$. The FK model tracks $c_a$.  In the HFK model we also track $\mathbf{c}_n(t) \in\mathbb{R}^N$, the per region degree of \emph{normal} tau. To describe the spreading between vertices in graph $\mathcal{G}$, we use the graph Laplacian matrix. We design an anisotropic diffusion partial differential equation (PDE) in brain domain $\mathcal{B}\subset\mathbb{R}^3$ to characterize the structural connectivity strength between ROIs. Then we construct the graph Laplacian based on the computed  structural connectivity. Given the graph Laplacian, we describe the FK and HFK model on the graph. Then we invert for the IC and model parameters. 

Before proceeding, we will like to clarify the proposed mathematical model is inspired by the biophysics of tau spreading  but it is not  biophysically accurate. That is, the tau abnormality is a measure of statistical discrepancy and not a biophysical quantity, for example concentration. It should be viewed as a phenomenological model that attempts to explain the dynamics of tau propagation using a minimal set of free parameters.

\subsection{Mathematical Models}
\label{subsec: model}

\subsubsection{Graph Laplacian}
\label{subsubsection: laplacian}
Let $\mathbf{x}\in\mathcal{B}$ denote a point in the brain. We denote $\mathbf{c}(\mathbf{x}, t)\in\left[0, 1\right]$ as a pseudo tau concentration at time $t$. We denote the diffusion source gray matter ROI that distributes concentration by $\mathcal{S}\subset\mathbb{R}^3$ and denote a target gray matter ROI that receive concentration as $\mathcal{T}\subset\mathbb{R}^3$. We use the concentration diffusion from $\mathcal{S}$ to $\mathcal{T}$ as a measure of the connectivity between $\mathcal{S}$ and $\mathcal{T}$. The diffusion process is modeled  by the following PDE:
\begin{subequations}
\label{eq: diffusion-pde}
\begin{align}
  \partial_t \mathbf{c} & = \nabla \cdot (\mathbf{K}(\mathbf{x}) \nabla \mathbf{c})\text{,} \quad t\in (0,T]\text{,}\\ 
  \mathbf{c}(\mathbf{x}, 0) & = \begin{cases}
    1 & \text{if } \mathbf{x}\in\mathcal{S} \\ 
    0 & \text{if } \mathbf{x}\in\mathcal{T} \text{,} 
  \end{cases} \\
  \frac{\partial \mathbf{c}}{\partial n} & = 0 \text{ on } \partial\mathcal{B} \text{,}
\end{align}
\end{subequations}
where $\mathbf{K}(\mathbf{x})$ is defined as
\begin{align}
  \mathbf{K}(\mathbf{x}) = \mathbf{D}(\mathbf{x}) (\mathbf{m}_{\text{wm}} + \Tilde{\alpha} \mathbf{m}_{\text{gm}}).
\end{align}
$\mathbf{D}(\mathbf{x})$ is the Diffusion Tensor Image (DTI) \cite{le2001diffusion}, $\mathbf{m}_{\text{wm}}$ and $\mathbf{m}_{\text{gm}}$ are the segmentation mask of white matter and grey matter, and $\Tilde{\alpha}\in\mathbb{R}_+$ is the ratio between the diffusivity in the white matter over the gray matter. From the literature, we set $\Tilde{\alpha}=10^{-2}$ \cite{giese1996migration}. We solve the above PDE in the time period $t\in\left[0, T\right]$, where $T$ is the time horizon. (We have normalized the brain domain to $\mathcal{B}=\left[0, 1\right]^3$.) 

The construction of the graph Laplacian involves the following steps: 1. For each source ROI, we calculate the solution to the proposed PDE \cref{eq: diffusion-pde}. This procedure is iteratively performed for all brain regions. 2. The integral concentration within each target ROI is computed as
\begin{equation}
W_{u, ij} = \int_{0}^T\int_{\mathcal{T}_j} \mathbf{c}(\mathbf{x}, t)\mathbf{1}_{\left\{\mathbf{c}>\mathbf{c}_{\infty}\right\}}d\mathbf{x}dt.
\end{equation}
Here, $\mathcal{T}_j$ represents the volume region of the $j^{\text{th}}$ ROI. $W_{u, ij}$ is the connectivity strength between the $i^{\text{th}}$ source $\mathcal{S}$ ROI and $j^{\text{th}}$ target ROI, and $\mathbf{c}_{\infty}={1}/{\left|\mathcal{B}\right|}\int_{\mathcal{S}}\mathbf{c}(\mathbf{x}, 0)d\mathbf{x}$ is the steady state  ($T=\infty$) solution of PDE for each voxel value. $W_{u, ij}$ contributes to the off-diagonal weight in the adjacency matrix to represent connectivity strength between vertices $i$ and $j$. We set the diagonal weight $W_{u, ii} = \sum_{j\neq i} W_{u, ij}$. 3. We normalize $\mathbf{W}$ row-wise dividing its diagonal entry. Therefore, the graph adjacency matrix is $\mathbf{W} = (\mathbf{W}_u + \mathbf{W}_u^\intercal) / 2$. 4. The graph Laplacian is then calculated as $\mathbf{L} = \text{diag}\left(\sum_{j=1}^N W_{ij}\right) - \mathbf{W}$, following the approach described in \cite{chung1997spectral}.

\subsubsection{Single-species FK Model}
\label{subsubsection: FK}
The Fisher-Kolmogorov model \cite{fisher1937wave,kolmogorov1937study}, also known as epidemic spreading model \cite{vogel2020spread}, is a popular temporal model to describe prion propagation. In this setup, the FK model is given by
\begin{subequations}
\label{eq: single-species forward}
\begin{align}
\frac{\partial \mathbf{c}_a}{\partial t} &= - \kappa \mathbf{L} \mathbf{c}_a + \rho \mathbf{c}_a \odot (1-\mathbf{c}_a) - \gamma \mathbf{c}_a\text{,}\\
\mathbf{c}_{a}(0) &= \mathbf{p}_0.
\end{align}
\end{subequations}
FK model involves three terms: diffusion, reaction and clearance. A diffusion term is defined by $- \kappa \mathbf{L} \mathbf{c}_a$, and $\kappa\in\mathbb{R}^+$ describes the migration rate of the model. A reaction term is defined by $\rho \mathbf{c}_a \odot (1 - \mathbf{c}_a)$ where $\odot$ stands for Hadamard vector product, and $\rho\in\mathbb{R}^+$ is the proliferation coefficient. We define a clearance term as $-\gamma \mathbf{c}_a$ and $\gamma \in\mathbb{R}^+$ is a clearance coefficient. $\mathbf{p}_{0}$ is the parametrization of $\mathbf{c}_a(0)$ described as tau initial condition. The FK model is widely used in describing tau propagation. However, in the FK model the high-tau regions in the IC remain high-tau throughout the simulation; no decay is possible. Therefore the model can not capture the dynamics of AD data when high concentrations are not present at the initial seeding location \cite{subramanian2020did,wen2023two}. The two-species model HFK below, addresses this shortcoming of FK.

\subsubsection{Two-species HFK Model} 
\label{subsubsection: HFK}
In the two-species model, we denote the normal tau and abnormal tau protein as $\mathbf{c}_n(t)$ and $\mathbf{c}_a(t)$, respectively, as following:
\begin{subequations}
\label{eq: two-species-forward}
\begin{align}
\frac{\partial \mathbf{c}_a}{\partial t} &= - \kappa \mathbf{L} \mathbf{c}_a + \rho \mathbf{c}_a \odot \mathbf{c}_n - \gamma \mathbf{c}_a\text{,}\\ 
\frac{\partial \mathbf{c}_n}{\partial t} &= - \rho \mathbf{c}_a \odot \mathbf{c}_n\text{,}\\
\mathbf{c}_{a}(0) &= \mathbf{p}_{0},\ \mathbf{c}_{n}(0) = \mathbf{1} - \mathbf{p}_{0}.
\end{align}
\end{subequations}
Here $\kappa, \rho, \gamma\in\mathbb{R}^+$ and $\mathbf{p}_{0}$ are unknown model parameters. We assume there is no diffusion and clearance for normal tau. Notice that since $\frac{\partial {c}_n^i}{\partial t}<0$ at all times until $c_n^i=0$, the maximum location of initial $\mathbf{c}_a$ will decay. This minimal change allows the location of maximum tau to temporally change, without introducing any new model parameters. The ODE model is defined for $t\in(0, T]$, and we set time horizon $T=1$ to non-dimensionalize the analysis \cite{subramanian2020did}.

\subsection{Inverse Problem}
\label{subsection: Inverse}

A patient-specific HFK model requires four model parameters: the initial condition $\mathbf{p}_0$ (which is a vector), the diffusion coefficient $\kappa$, the reaction coefficient $\rho$, and the clearance coefficient $\gamma$.
Instead of fixing the initial condition to be located EC, which has been demonstrated the limited performance \cite{vogel2021four,wen2023two}, $\mathbf{p}_0$ needs to be solved. This is challenging as we use just a  single tau PET scan to drive the reconstruction. It can be shown that this is a highly ill-posed inverse problem and requires regularization.  In \cite{subramanian2020did,scheufele2020fully,subramanian2022ensemble}, we dealt with similar problems and demonstrated the need to add sparsity constraints to the inverse problem for three main reasons:
\begin{enumerate}
\item Without sparsity constraints, the estimated initial condition would be simply a scaling of the signal, as shown in our previous work on tumors \cite{subramanian2020did}.
\item The linearized diffusion-reaction system indicates the potential ill-posedness of our model \cite{subramanian2020did}, requiring extra constraints for the system.
\item We aim to investigate the source of Alzheimer's disease and verify the initial condition hypothesis, as in \cite{vogel2020spread,vogel2021four}. Sparsity allows to localize the initiation.
\end{enumerate}
Adapting a similar procedure, we impose $\ell_0$ sparsity constraints on the problem. For the HFK model, the inversion can be performed by defining the following optimization problem:
\begin{subequations}
\begin{align}
\min_{\mathbf{p}_0, \kappa, \rho, \gamma}&\ J \coloneqq \frac{1}{2}\left\|\mathbf{c}_{a}(T)-\mathbf{d}\right\|_2^2 + \beta_1\sum_{i} \log(1-[\mathbf{p}_0]_{i})\text{,}\\
s.t.&\begin{cases}
\ \frac{\partial \mathbf{c}_a}{\partial t} = - \kappa \mathbf{L} \mathbf{c}_a + \rho \mathbf{c}_a\odot \mathbf{c}_n - \gamma \mathbf{c}_a\\
\ \frac{\partial \mathbf{c}_n}{\partial t} = - \rho \mathbf{c}_a\odot \mathbf{c}_n\\
\ \mathbf{c}_{a}(0) = \mathbf{p}_{0}\\
\ \mathbf{c}_{n}(0) = \mathbf{1} - \mathbf{p}_{0}\\
\ \kappa,\ \rho,\ \gamma \geq 0\\
\ \left\|\mathbf{p}_0\right\|_0 \leq s^{\text{max}}\text{.}
\end{cases}
\end{align}
\end{subequations}
Here, $s^{\text{max}}\in\mathbb{N}_+$ is the maximum number of regions allowed to be activated (nonzero) in IC (manually set), and $\beta_1$ is a hyperparameter. We minimize the residual between the simulated abnormal tau $\mathbf{c}_a$ and observation data $\mathbf{d}$ in the $\ell_2$ norm. The regularization proves essential for ensuring inversion stability, particularly in the face of the highly heterogeneous AD data. The optimization is subject to the ordinary differential equations that describe the propagation of normal/abnormal tau and bound constraints for the parameters.

Next we derive adjoint equations that are used to compute derivatives. The derivation is based on defining a Lagrangian functional, adjoint variables, and taking variations. 
We define $\bm{\alpha}_a, \ \bm{\alpha}_n\in\mathbb{R}^N$ as adjoint variables of $\mathbf{c}_a$ and $\mathbf{c}_n$ separately. Satisfying the first-order optimality with respect to the Lagrangian multipliers results in the forward model as shown in \cref{eq: two-species-forward}, and optimality with respect to $\mathbf{c}_a$ and $\mathbf{c}_n$ leads to the adjoint equations as follows:
\begin{subequations}
\begin{align}
\frac{\partial \bm{\alpha}_a}{\partial t} &= \kappa \mathbf{L}^\intercal \bm{\alpha}_a + \rho \mathbf{c}_n\odot (\bm{\alpha}_n - \bm{\alpha}_a) + \gamma \bm{\alpha}_a\text{,}\\
\frac{\partial \bm{\alpha}_n}{\partial t} &= \rho \mathbf{c}_a\odot (\bm{\alpha}_n - \bm{\alpha}_a)\text{,}\\
\bm{\alpha}_{a}(T) &=\mathbf{d} - \mathbf{c}_{a}(T)\text{,} \\
\bm{\alpha}_{n}(T) &= \mathbf{0}\text{.}
\end{align}
\label{eq:adjoint}
\end{subequations}
Using the adjoints, the  gradients of $\mathcal{J}$ with respect to the unknowns $\mathbf{p}_{0}$, $\kappa$, $\rho$ and $\gamma$ are computed by 
\begin{subequations}
\label{eq: gradient descent}
\begin{align}
\frac{\partial \mathcal{J}}{\partial \mathbf{p}_0} &= - \bm{\alpha}_{a}(0) + \bm{\alpha}_{n}(0) + \beta_1{\mathbf{1}}_{\left\{j: p_{0}^j \neq 0\right\}}\oslash(\mathbf{p}_0 - \mathbf{1})\text{,}\\
\frac{\partial \mathcal{J}}{\partial \kappa} &= \int_0^T  \bm{\alpha}_a^\intercal (\mathbf{L} \mathbf{c}_a) dt\text{,}\\
\frac{\partial \mathcal{J}}{\partial \rho} &= \int_0^T (\bm{\alpha}_n - \bm{\alpha}_a)^\intercal \mathbf{c}_a \odot \mathbf{c}_n dt\text{,}\\
\frac{\partial \mathcal{J}}{\partial \gamma} &= \int_0^T\bm{\alpha}_a^\intercal \mathbf{c}_a dt\text{,}
\end{align}
\end{subequations}
where $\oslash$ is element-wise division, and vector ${\mathbf{1}}_{\left\{j: p_{0}^j \neq 0\right\}}$ equals one for the $j^{\text{th}}$ entry if $j\in\left\{j: p_{0}^j \neq 0\right\}$. In order to evaluate the gradients, we first solve the forward problem \cref{eq: two-species-forward}, next we solve the adjoint \cref{eq:adjoint} and finally we evaluate the gradients w.r.t. the parameters. To update the parameters, for each inversion, we use a quasi-newton L-BFGS method given gradients in \cref{eq: gradient descent} \cite{zhu1997algorithm}. In the following, we discuss how to enforce the sparsity constraint for IC.

\subsubsection{IC Inversion algorithm}
\label{subsubsection: IC inversion} 
As discussed, we impose an $\ell_0$ sparsity constraint on $\mathbf{p}_0$. Our solver is based on the original scheme for linear least squares \cite{tropp2010computational}, and its extension  to nonlinear least squares in \cite{subramanian2020did}. We define $\text{supp}(\mathbf{p})=\left\{j: p_j\neq0\right\}$, and $\left.\kern-\nulldelimiterspace\mathbf{p}\right\vert^s=\begin{cases} p_j, & j\in {S},\\ 0, & \text{otherwise,} \end{cases}$ where ${S}$ represents the set of indices corresponding to the $s$ largest components of $\mathbf{p}$. Also, we define two additional hyperparameters $\beta_2$ and $\eta$. The hyperparameters $\beta_1$ and $\beta_2$ are chosen to ensure a balance between achieving a low relative error and maintaining low sensitivity of inversion when observations are perturbed. Specifically, we initialize $\beta_1$ at 100 and reduce it by a factor of 0.1 every time the sparse initial condition $\mathbf{p}_0$ is updated. The value of $\beta_2$ is set to 0.5. For the objective value tolerance, we use $\eta = \num{1e-8}$. We set the initial guess $\kappa^0=0,\ \rho^0=0,\ \gamma^0=0,\ \mathbf{p}_0^0=\mathbf{0}$ and we observe limited sensitivity of final residual to the initial guess. We use the LSODA as forward ODE solver \cite{petzold1983automatic} and store the computed solution with time step size $\num{1E-4}$. The quit condition of optimizer L-BFGS uses gradient tolerance as $\num{1e-3}$. We setup the bound constraints for the parameters $\kappa,\ \rho,\ \gamma\in\left[0, 20\right]$ to improve the speed of the optimizer. The scheme is  summarized in \Cref{algo:ic-inv}. 

\begin{algorithm}[!t]
\footnotesize
\caption{{Inversion algorithm}}
\label{algo:ic-inv}
\begin{algorithmic}[1]
\State $\mathbf{p}_0^0=\mathbf{0},\ \kappa^0=0,\ \rho^0=0,\ \gamma^0=0,\ i=0,\ m=N$
\While {$m>2s^{\text{max}}$}
	\State $i\gets i+1$
	\State $m=\max(\beta_2m, 2s^{\text{max}})$
	\State $\left(\mathbf{p}_0^i, \kappa^i, \rho^i, \gamma^i\right)$ are updated by \Cref{eq: gradient descent} without sparsity constraint
	\State $S\gets \text{supp}(\left.\kern-\nulldelimiterspace\mathbf{p}_0^i\right\vert^m)$
	\State $\left(\mathbf{p}_0^i, \kappa^i, \rho^i, \gamma^i\right)$ are updated by \Cref{eq: gradient descent} in subspace $S$
\EndWhile

\While {$\left|\mathcal{J}^{i+1} - \mathcal{J}^{i}\right|<\eta$}
	\State $i\gets i+1$
	\State compute $\frac{\partial \mathcal{J}}{\partial \mathbf{p}_0}$ of $\mathbf{p}_0^i$ from \Cref{eq: gradient descent}
	\State $S\gets \text{supp}(\mathbf{p}_0^i) \bigcup \text{supp}\left(\left.\kern-\nulldelimiterspace\frac{\partial \mathcal{J}}{\partial \mathbf{p}_0}\right\vert^{s^{\text{max}}}\right)$
	\State $\left(\mathbf{p}_0^i, \kappa^i, \rho^i, \gamma^i\right)$ are updated by \Cref{eq: gradient descent} in subspace $S$
	\State $\mathbf{p}_0^i\gets \left.\kern-\nulldelimiterspace\mathbf{p}_0^i\right\vert^{s^{\text{max}}}$
\EndWhile
\State $S\gets \text{supp}(\mathbf{p}_0^i)$
\State $\left(\mathbf{p}_0^*, \kappa^*, \rho^*, \gamma^*\right)$ is updated by \Cref{eq: gradient descent} in subspace $S$
\end{algorithmic}
\end{algorithm}

\subsection{Data, Workflow, and Compute Platforms}
\label{subsection: workflow}
\subsubsection{Data }
\label{subsubsection: data}

We use preprocessed MRI and PET images obtained from the ADNI database,
which includes 455 CN, 212 MCI, and 45 AD subjects. Among these, 365 participants are female and 347 are male, with an average ($\pm$standard deviation) age of 73.85 ($\pm$8.08) years, acquired between September 30, 2015 and July 28, 2022. We use MRI scans for brain registration and normalization. The most recent PET image of each patient is utilized in the inversion algorithm. If a patient has multiple PET images, the second most recent PET image is used for validation purposes, as detailed in \Cref{ss:clinical_test}. We generate synthetic data for inversion algorithm analysis using the proposed HFK model, a set of specific model parameters, and EC as IC.
We also use 20 DTI scans from normal subjects  from the Harvard Aging Brain Study (HABS) \cite{dagley2017harvard} to generate tractography information used to generate the graph Laplacian.

\subsubsection{Preprocessing}
\label{subsubsection: processing}
We use the MRI scans to define the graph Laplacian and normalize the data to a common atlas; and use the PET scans to drive the analysis.   We follow the preprocessing pipeline described in \cite{vogel2020spread}. For each subject, the T1 MRI scan is affinely registered to a brain template using FSL \cite{smith2004advances}. Tau-PET data correspond to time-averaged  mean and standard deviation  over 5-minute image frames. The MRI and PET scans are parcellated using the MUSE template~\cite{doshi2016muse}, which is mapped from normal brain template space to patient space using ANTs software \cite{avants2009}. 

The $d$ values are defined as follows. We compare the tau distribution of a ROI with the tau distribution in the reference cerebellum. We employ the MMD metric \cite{gretton2012kernel} to quantify the  distance  between these two distributions. The MMD score, denoted as $\mu_i = \text{MMD}(p_i, q) \in [0, \infty]$, measures the distance  between the distribution $p_i$ from the $i^{\text{th}}$ ROI and $q$ from the same subject cerebellum. A high $\mu_i$ indicates a high tau abnormality in each parcel. Additionally, we normalize $\mu_i$ by $d^i=1 - \exp({-\mu_i  \sigma})$, where $\sigma$ is a hyperparameter; we have select $\sigma = 0.3$ by maximize the average fitting performance in our subject population.

Due to the limited availability of DTI for ADNI patients, we used HBAS normal subject data  to generate 20 connectivity matrices $\mathbf{W}$.
Once constructed, we average these graph adjacency matrices into a single $\mathbf{W}$, which we then use in all of our experiments. In other words, the graph Laplacian is not subject-specific. 

Finally, all the runs took place on the Frontera and Lonestar6 system at the Texas Advanced Computing Center (TACC) at The University of Texas at Austin for all the experiments.  The solver is written in Python 3.9.

%% file: results.tex
\section{Results}
\label{sec:results}

\setlength{\aboverulesep}{0pt}
\setlength{\belowrulesep}{0pt}
\begin{table*}[!t]
\caption{Sensitivity of inversion IC as a function of the time horizon, assessed with synthetic data: We select ground truth  $(\kappa^*,\ \rho^*,\ \gamma^*)$, generate synthetic data, and then  reconstruct using observations from different time horizons. The time horizon is adjusted by changing the initial time $t_0$ at which we start the simulation. We observe data at $t=1$. Therefore,  the time  horizon of the simulation is $1-t_0$ and the larger $t_0$ the smaller the time horizon.  We report the relative errors $e_{\bold{d}}$, $e_{\kappa}$, $e_{\rho}$ and $e_{\gamma}$. We also provide the $r^2$. We bold the numbers that performs best in each synthetic case. We repeated this experiment 100 times and report the mean and standard deviation. The results suggest that inverting for the IC at $t_0=0$ performs the best across all cases.}
\begin{adjustbox}{width=\textwidth,center}
\small \setlength{\tabcolsep}{1pt}
\begin{tabular}{c|c|c|c|c|c|c}
\toprule
\text{Parameter ($\kappa^*, \rho^*, \gamma^*$)}  & $\text{IC timepoint } t_0$ & $e_{\bold{d}}$ & $r^2$ & $e_{\kappa}$ & $e_{\rho}$ & $e_{\gamma}$ \\
\midrule
\rowcolor[gray]{0.8}
$(4, 5, 1)$ & $\begin{tabular}{c} $0$\\ $0.95$\\ $0.99$ \end{tabular}$ & $\begin{tabular}{c} $\num[math-rm=\mathbf]{1.01E-1} \pm \num[math-rm=\mathbf]{2.65E-3}$ \\ $\num{2.51e-1} \pm \num{1.88e-2}$ \\ $\num{4.02e-1} \pm \num{1.71e-2}$ \end{tabular}$ & $ \begin{tabular}{c} $\num[math-rm=\mathbf]{9.22E-1} \pm \num[math-rm=\mathbf]{4.55E-3} $\\ $ \num{5.25e-1} \pm \num{7.07e-2}$ \\ $\num{-2.08e-1} \pm \num{1.17e-2}$ \end{tabular}$ & $\begin{tabular}{c} $\num[math-rm=\mathbf]{8.83E-2} \pm \num[math-rm=\mathbf]{3.57E-2} $\\ $\num{4.00} \pm \num{0}$ \\ $\num{4.00} \pm \num{0}$ \end{tabular}$ &  $\begin{tabular}{c} $\num[math-rm=\mathbf]{1.05E-1} \pm \num[math-rm=\mathbf]{3.54E-2}$ \\ $\num{1.90} \pm \num{1.18}$\\ $\num{1.19} \pm \num{8.25e-1}$ \end{tabular}$ & $\begin{tabular}{c} $ \num[math-rm=\mathbf]{3.39E-1} \pm \num[math-rm=\mathbf]{1.22E-1}$ \\ $\num{7.28} \pm \num{4.09}$ \\ $\num{1.88e1} \pm \num{6.50e-1}$ \end{tabular}$ \\ \midrule 
$(4, 4, 1)$ & $\begin{tabular}{c} $0$\\ $0.95$\\ $0.99$ \end{tabular}$ & $\begin{tabular}{c} $\num[math-rm=\mathbf]{1.00e-1} \pm \num[math-rm=\mathbf]{2.10e-3}$ \\ $\num{3.69e-1} \pm \num{3.76e-2}$ \\ $\num{6.05e-1} \pm \num{2.46e-2}$ \end{tabular}$ & $ \begin{tabular}{c} $\num[math-rm=\mathbf]{9.46e-1} \pm \num[math-rm=\mathbf]{1.20e-3} $\\ $ \num{2.64e-1} \pm \num{1.58e-1}$ \\ $\num{-9.55e-1} \pm \num{1.77e-1}$ \end{tabular}$ & $\begin{tabular}{c} $\num[math-rm=\mathbf]{6.33e-2} \pm \num[math-rm=\mathbf]{1.49e-2} $\\ $\num{4.00} \pm \num{0}$ \\ $\num{4.00} \pm \num{0}$ \end{tabular}$ &  $\begin{tabular}{c} $\num[math-rm=\mathbf]{8.66e-2} \pm \num[math-rm=\mathbf]{1.88e-2}$ \\ $\num{1.86} \pm \num{1.62}$\\ $\num{1.00} \pm \num{0}$ \end{tabular}$ & $\begin{tabular}{c} $ \num[math-rm=\mathbf]{2.85e-1} \pm \num[math-rm=\mathbf]{8.80e-2}$ \\ $\num{1.02e1} \pm \num{5.61}$ \\ $\num{1.90e1} \pm \num{0}$ \end{tabular}$ \\ \midrule 
\rowcolor[gray]{0.8}
$(6, 5, 1)$ & $\begin{tabular}{c} $0$\\ $0.95$\\ $0.99$ \end{tabular}$ & $\begin{tabular}{c} $\num[math-rm=\mathbf]{1.08e-1} \pm \num[math-rm=\mathbf]{3.81e-2}$ \\ $\num{2.18e-1} \pm \num{1.78e-2}$ \\ $\num{3.84e-1} \pm \num{1.58e-2}$ \end{tabular}$ & $ \begin{tabular}{c} $\num[math-rm=\mathbf]{7.93e-1} \pm \num[math-rm=\mathbf]{2.45e-1} $\\ $ \num{2.41e-1} \pm \num{1.34e-1}$ \\ $\num{-1.34} \pm \num{2.43e-1}$ \end{tabular}$ & $\begin{tabular}{c} $\num[math-rm=\mathbf]{5.69e-2} \pm \num[math-rm=\mathbf]{6.61e-2} $\\ $\num{2.33} \pm \num{4.44e-16}$ \\ $\num{2.33} \pm \num{4.44e-16}$ \end{tabular}$ &  $\begin{tabular}{c} $\num[math-rm=\mathbf]{6.84e-2} \pm \num[math-rm=\mathbf]{6.31e-1}$ \\ $\num{2.40} \pm \num{9,34e-1}$\\ $\num{1.43} \pm \num{8.98e-1}$ \end{tabular}$ & $\begin{tabular}{c} $ \num[math-rm=\mathbf]{2.55e-1} \pm \num[math-rm=\mathbf]{2.25e-1}$ \\ $\num{9.20} \pm \num{3.40}$ \\ $\num{1.88e1} \pm \num{6.22e-1}$ \end{tabular}$ \\ \midrule 
$(4, 5, 1.5)$ & $\begin{tabular}{c} $0$\\ $0.95$\\ $0.99$ \end{tabular}$ & $\begin{tabular}{c} $\num[math-rm=\mathbf]{1.03e-1} \pm \num[math-rm=\mathbf]{4.83e-3}$ \\ $\num{3.25e-1} \pm \num{2.85e-2}$ \\ $\num{5.32e-1} \pm \num{2.38e-2}$ \end{tabular}$ & $ \begin{tabular}{c} $\num[math-rm=\mathbf]{9.26e-1} \pm \num[math-rm=\mathbf]{6.83e-3} $\\ $ \num{2.85e-1} \pm \num{1.18e-1}$ \\ $\num{-9.02e-1} \pm \num{1.88e-1}$ \end{tabular}$ & $\begin{tabular}{c} $\num[math-rm=\mathbf]{1.64e-1} \pm \num[math-rm=\mathbf]{6.61e-2} $\\ $\num{4.00} \pm \num{0}$ \\ $\num{4.00} \pm \num{0}$ \end{tabular}$ &  $\begin{tabular}{c} $\num[math-rm=\mathbf]{2.06e-1} \pm \num[math-rm=\mathbf]{3.37e-2}$ \\ $\num{1.84} \pm \num{1.19}$\\ $\num{9.98e-1} \pm \num{1.27e-2}$ \end{tabular}$ & $\begin{tabular}{c} $ \num[math-rm=\mathbf]{4.90e-1} \pm \num[math-rm=\mathbf]{8.33e-2}$ \\ $\num{6.66} \pm \num{3.26}$ \\ $\num{1.23e1} \pm \num{3.55e-15}$ \end{tabular}$ \\ \bottomrule
\end{tabular}
\end{adjustbox}
\label{tab:syn_time}
\end{table*}

\setlength{\aboverulesep}{0pt}
\setlength{\belowrulesep}{0pt}
\begin{table*}[!t]
\caption{Accuracy of the inversion algorithm in the presence of noise but assuming the model is correct. Assessed using synthetic data. Using four ground truth cases we test the accuracy in the presence of Gaussian noise levels of $0\%, 5\%$, and $10\%$ to MMD score. We report the relative errors $e_{\bold{d}}$ and $e_{\bold{p}_0}$ respectively. The ground truth values for each model coefficient $\iota$ are represented as $\iota^*$, with their respective relative errors reported as $e_{\iota}$. We repeat each test 100 times, and we report the mean and standard deviation of errors across all metrics.}
\begin{adjustbox}{width=\textwidth,center}
\small \setlength{\tabcolsep}{1pt}
\begin{tabular}{c|c|c|c|c|c|c|c}
\toprule
\text{Parameter ($\kappa^*, \rho^*, \gamma^*$)} & $\text{Noise level}$ & $e_{\bold{d}}$ & $r^2$ & $e_{\kappa}$ & $e_{\rho}$ & $e_{\gamma}$ & $e_{\bold{p}_0}$ \\
\midrule
\rowcolor[gray]{0.8}
$(4, 5, 1)$ & $\begin{tabular}{c} $0$\\ $5\%$\\ $10\%$ \end{tabular}$ & $\begin{tabular}{c} $\num{2.26e-6}$\\ $\num{6.31e-2} \pm \num{7.38e-3}$ \\ $\num{9.71e-2}\pm\num{9.86e-4}$ \end{tabular}$ & $\begin{tabular}{c} $\num{1.00}$ \\ $\num{9.83e-1}\pm\num{2.04e-2}$\\ $\num{9.26e-1}\pm\num{2.98e-3}$ \end{tabular}$ &  $\begin{tabular}{c} $\num{1.21e-5}$\\ $\num{1.72e-2}\pm\num{2.04e-3}$\\ $\num{5.65e-2}\pm\num{3.29e-3}$ \end{tabular}$ & $\begin{tabular}{c} $\num{1.10e-5}$\\ $\num{1.65e-2}\pm\num{1.08e-3}$\\ $\num{8.21e-2}\pm\num{7.36e-3}$ \end{tabular}$ & $\begin{tabular}{c} $\num{3.62e-5}$\\ $\num{1.04e-1}\pm\num{2.28e-2}$\\ $\num{2.43e-1}\pm\num{1.79e-2}$ \end{tabular}$ & $\begin{tabular}{c} $\num{2.77e-11}$\\ $\num{9.90e-2}\pm\num{1.19e-3}$\\ $\num{1.30e-1}\pm\num{1.64e-2}$ \end{tabular}$ \\ 
$(3, 5, 1)$ & $\begin{tabular}{c} $0$\\ $5\%$\\ $10\%$ \end{tabular}$ & $\begin{tabular}{c} $\num{2.73e-5}$\\ $\num{4.89e-2}\pm\num{3.05e-4}$ \\ $\num{9.81e-2}\pm\num{9.45e-4}$ \end{tabular}$ & $\begin{tabular}{c} $\num{1.00}$ \\ $\num{9.88e-1}\pm\num{3.42e-4}$\\ $\num{9.55e-1}\pm\num{2.80e-3}$ \end{tabular}$ &  $\begin{tabular}{c} $\num{1.35e-5}$\\ $\num{2.26e-2}\pm\num{1.36e-3}$\\ $\num{3.63e-2}\pm\num{3.60e-3}$ \end{tabular}$ & $\begin{tabular}{c} $\num{9.76e-5}$\\ $\num{3.06e-2}\pm\num{2.54e-3}$\\ $\num{5.60e-2}\pm\num{3.26e-3}$ \end{tabular}$ & $\begin{tabular}{c} $\num{2.90e-4}$\\ $\num{8.09e-2}\pm\num{7.02e-3}$\\ $\num{1.59e-1}\pm\num{9.80e-2}$ \end{tabular}$ & $\begin{tabular}{c} $\num{1.07e-4}$\\ $\num{5.85e-2}\pm\num{5.91e-3}$\\ $\num{7.13e-2}\pm\num{8.07e-3}$ \end{tabular}$ \\  
\rowcolor[gray]{0.8}
$(4, 4, 1)$ & $\begin{tabular}{c} $0$\\ $5\%$\\ $10\%$ \end{tabular}$ & $\begin{tabular}{c} $\num{8.44e-5}$\\ $\num{5.58e-2}\pm\num{2.13e-3}$ \\ $\num{9.75e-2}\pm\num{1.75e-3}$ \end{tabular}$ & $\begin{tabular}{c} $\num{1.00}$ \\ $\num{9.80e-1}\pm\num{1.97e-2}$\\ $\num{9.40e-1}\pm\num{1.05e-3}$ \end{tabular}$ &  $\begin{tabular}{c} $\num{1.37e-4}$\\ $\num{2.13e-2}\pm\num{3.45e-3}$\\ $\num{2.72e-2}\pm\num{2.68e-3}$ \end{tabular}$ & $\begin{tabular}{c} $\num{1.28e-3}$\\ $\num{4.77e-2}\pm\num{2.61e-3}$\\ $\num{6.97e-2}\pm\num{3.91e-3}$ \end{tabular}$ & $\begin{tabular}{c} $\num{1.86e-3}$\\ $\num{1.27e-1}\pm\num{9.67e-2}$\\ $\num{1.67e-1}\pm\num{1.01e-2}$ \end{tabular}$ & $\begin{tabular}{c} $\num{3.91e-3}$\\ $\num{1.55e-2}\pm\num{2.78e-3}$\\ $\num{6.75e-2}\pm\num{8.87e-3}$ \end{tabular}$ \\ 
$(4, 5, 0.5)$ & $\begin{tabular}{c} $0$\\ $5\%$\\ $10\%$ \end{tabular}$ & $\begin{tabular}{c} $\num{6.10e-4}$\\ $\num{5.60e-2}\pm\num{7.37e-3}$ \\ $\num{1.13e-1}\pm\num{4.12e-2}$ \end{tabular}$ & $\begin{tabular}{c} $\num{1.00}$ \\ $\num{9.81e-1}\pm\num{5.56e-3}$\\ $\num{8.91e-1}\pm\num{1.43e-2}$ \end{tabular}$ &  $\begin{tabular}{c} $\num{7.57e-4}$\\ $\num{3.93e-2}\pm\num{3.56e-3}$\\ $\num{8.65e-2}\pm\num{1.45e-2}$ \end{tabular}$ & $\begin{tabular}{c} $\num{7.25e-3}$\\ $\num{1.25e-1}\pm\num{1.77e-2}$\\ $\num{1.28e-1}\pm\num{2.29e-2}$ \end{tabular}$ & $\begin{tabular}{c} $\num{2.06e-2}$\\ $\num{6.05e-1}\pm\num{6.16e-2}$\\ $\num{6.39e-1}\pm\num{1.23e-1}$ \end{tabular}$ & $\begin{tabular}{c} $\num{3.19e-2}$\\ $\num{2.14e-1}\pm\num{4.54e-2}$\\ $\num{2.45e-1}\pm\num{4.46e-2}$ \end{tabular}$ \\ \bottomrule
\end{tabular}
\end{adjustbox}
\label{tab:syn_invert_p0}
\end{table*}

\setlength{\aboverulesep}{0pt}
\setlength{\belowrulesep}{0pt}
\begin{table}[!t]
\caption{Classification results for SUVR and MMD features. The table shows the classification accuracy for both training and test case using SUVR, tau probability or MMD as features. As we can observe, the MMD exhibits significantly higher accuracy compared to the other two methods.}
\begin{adjustbox}{width=1\linewidth,center}
\small \setlength{\tabcolsep}{1pt}
\begin{tabular}{c|ccc|>{\columncolor[gray]{0.8}}c>{\columncolor[gray]{0.8}}c>{\columncolor[gray]{0.8}}c}
\toprule
\text{Feature}  & \multicolumn{3}{c}{\text{Train}} & \multicolumn{3}{c}{\text{Test}}\\
\cmidrule{2-7} 
 & \text{CN} & \text{AD} &  \text{All} & \text{CN} & \text{AD} & \text{All} \\ \midrule
\text{SUVR} &  \num{9.76e-1}& \num{8.14e-1} & \num{9.13e-1} & \num{9.31e-1} & \num{8.33e-1} & \num{8.94e-1} \\ 
\text{Tau Probability\cite{vogel2020spread}} &  \num{9.52e-1}& \num{8.51e-1} & \num{9.14e-1} & \num{9.32e-1} & \num{8.89e-1} & \num{9.15e-1} \\
\text{MMD} &  \num{1.00}& \num{9.26e-1} & \num{9.71e-1} & \num[math-rm=\mathbf]{9.66e-1} & \num[math-rm=\mathbf]{1.00} & \num[math-rm=\mathbf]{9.79e-1} \\  \bottomrule
\end{tabular}
\end{adjustbox}
\label{tab:classification}
\end{table}

\subsection{Analysis of Inversion Algorithm Using Synthetic data.}
We use synthetic data to address two questions:
We analyze our model and inversion algorithm starting from synthetic tests by answering two questions.
\begin{enumerate}[label=(\textbf{SQ\arabic*}), itemindent=*,leftmargin=23pt]
\item Since many ADNI subjects have longitudinal PET data, why not using the first scan as an initial condition and then fit to the second scan? Why do we need invert for an initial condition from a single scan?
\item How accurate are the reconstructions of our inversion algorithm assuming no modeling errors but noisy data?
\end{enumerate}
In a nutshell, the answer to the \textbf{(SQ1)} is that the time interval between PET acquisition is too short, in disease timescales, and the inversion becomes unstable. This is a short time horizon instability since the signal is not different enough to inform the dynamics. The answer to \textbf{(SQ2)} is that our solver near machine precision accurate, in the absence of noise; but in the presence of noise and for one parameter in particular ($\gamma$) the error can be up to 20\%. We give details below.

\par\medskip
\noindent\textbf{(SQ1)} \textit{Effect of time horizon: } The synthetic data is generated using the HFK model, we set $\bold{p}_0$ to be one at  EC and zero elsewhere at $t=0$. We run the forward model with four different parameter combinations as shown in \Cref{tab:syn_time}. For each parameter combination, we run a reconstruction fitting our model to the system solution at $t=1.0$ and use time horizon $1-t_0$, where $t_0=0,\ 0.95$, and $0.99$ respectively. Notice that in this experiment \emph{we use data from two time points, $t_0$ and $t=1$} and we just invert for the three scalar coefficients  not the IC, which is given at the observation $t_0$. We introduce $10\%$ Gaussian noise to both initial and end scans. \Cref{tab:syn_time} summarizes the results.

For each parameter set, we repeat the experiment 100 times and compute mean and standard deviation of several error metrics. We report $e_{\bold{d}} = \frac{\left\|\bold{c}_a(T) - \bold{d}\right\|_2}{\left\|\bold{d}\right\|_2}$, which is the relative error between the fitting result and the observation under the $\ell_2$ norm; and we report the $r^2$ score given by $r^2(\bold{d}, \bold{c}_a(T)) = 1 - \frac{\sum_{i}({d}_{i} - c_{a,i})^2}{\sum_{i}({d}_{i} - \bar{d})^2}\in\left[-\infty, 1\right]$, which indicates how much of the spatial pattern that the observation can be explained by the fitting result, where $\bar{d}$ is the mean of entries of $\bold{d}$. Good fits give $r^2$ near one; poor fits give  $r^2<0$.
For each model coefficient $\iota \in (\kappa, \rho, \gamma)$, we denote the corresponding ground truth parameter as $\iota^*$, and we report the relative error using $e_{\iota}$. 

We observe that the longer time horizon  (IC timepoint $t_0=0$) the more accurate the reconstruction with  smaller relative errors, and higher $r^2$ scores. But as the time horizon shrinks, as it is the case with real longitudinal data, the quality of the fitting deteriorates. The experiment emphasizes the scenario wherein using two snapshot data for parameter fitting may lead to mathematical ill-posedness during the inversion process. In our next experiment we use only one time snap shot and reconstruct for the scalar coefficients \emph{and} the IC at $t_0=0$.
\par\medskip

\noindent\textbf{(SQ2)} \textit{Synthetic inversion performance: }In this test, we use the HFK model to generate data at $t=1.0$ for specific combinations of parameters shown in \Cref{tab:syn_invert_p0}. We subsequently perform an inversion process using our proposed algorithm to estimate both IC and model coefficients. We investigate three cases: no noise, 5\% Gaussian noise, and 10\% Gaussian noise, which are added to the data. The numerical outcomes are summarized in \Cref{tab:syn_invert_p0}. We report the $\ell_2$ relative errors for IC, represented as $e_{\mathbf{p}_0}$. Notably, our inversion algorithm consistently demonstrates its effectiveness in reconstructing the observed data and accurately estimating the parameters, including the IC location.

\subsection{Clinical Data Analysis}
\label{ss:clinical_test}
We first present the performance of our algorithm for HFK model compared to FK model using clinical dataset. Then we demonstrate the model's sensitivity to templates, parcellations, and connectivity matrices. Finally, we analyze the model capability on longitudinal clinical data. We address the following questions:
\begin{enumerate}[label=(\textbf{CQ\arabic*}), itemindent=*, leftmargin=23pt]
\item How does the MMD abnormality  indicator compares to regional SUVR and tau positive probability as in \cite{vogel2020spread}?
\item How do the reconstructions using HFK and FK differ?
\item What is the performance of the models for the cohort ADNI dataset?
\item How sensitive is the inversion algorithm to templates, parcellations, and connectivity matrices?
\item How can we evaluate the model using longitudinal data given the results from the synthetic data experiments?
\end{enumerate}
\par\medskip

\setlength{\aboverulesep}{0pt}
\setlength{\belowrulesep}{0pt}
\begin{table*}[!t]
\caption{Inversion of four clinical subjects using different models. We compare the HFK and FK models in two different scenarios: IC ($s^\max$=5) and IC=EC. We report the relative errors $e_{\bold{d}}$, and $r^2$ score. We highlight the best performance in bold.}
\begin{adjustbox}{width=1\textwidth,center}
\small \setlength{\tabcolsep}{1pt}
\begin{tabular}{c|>{\columncolor[gray]{0.8}}c|>{\columncolor[gray]{0.8}}c|c|c|>{\columncolor[gray]{0.8}}c|>{\columncolor[gray]{0.8}}c|c|c}
\toprule
\text{SubjectId}  & \multicolumn{2}{c}{\text{HFK (IC $s^\max$=5)}} & \multicolumn{2}{c}{\text{FK (IC $s^\max$=5)}} & \multicolumn{2}{c}{\text{HFK (IC=EC)}} & \multicolumn{2}{c}{\text{FK (IC=EC)}} \\
\cmidrule{2-9} 
 & $r^2$ & $e_{\bold{d}}$ &  $r^2$ & $e_{\bold{d}}$ &  $r^2$ & $e_{\bold{d}}$ &  $r^2$ & $e_{\bold{d}}$ \\ \midrule
\text{032\_S\_6602} &  $\num[math-rm=\mathbf]{6.20e-1}\pm\num[math-rm=\mathbf]{4.82e-2}$ & $\num[math-rm=\mathbf]{3.23e-1}\pm\num[math-rm=\mathbf]{1.67e-2}$ & $\num{5.62e-1}\pm\num{3.19e-2}$ & $\num{3.96e-1}\pm\num{2.46e-2}$ & $\num{4.83e-1}\pm\num{7.34e-2}$ & $\num{4.02e-1}\pm\num{1.23e-2}$ & $\num{-4.80e-1}\pm\num{1.32e-3}$ & $\num{6.36e-1}\pm\num{1.11e-3}$ \\ 
\text{168\_S\_6828} &  $\num[math-rm=\mathbf]{4.76e-1}\pm\num[math-rm=\mathbf]{5.15e-2}$ & $\num[math-rm=\mathbf]{4.04e-1}\pm\num[math-rm=\mathbf]{2.54e-2}$ & $\num{3.12e-1}\pm\num{2.14e-2}$ & $\num{5.58e-1}\pm\num{2.70e-2}$ & $\num{3.78e-1}\pm\num{3.38e-2}$ & $\num{5.02e-1}\pm\num{7.94e-2}$ & $\num{2.16e-1}\pm\num{2.34e-2}$ & $\num{6.02e-1}\pm\num{4.23e-2}$ \\ 
\text{116\_S\_6100} &  $\num[math-rm=\mathbf]{4.14e-1}\pm\num[math-rm=\mathbf]{1.01e-2}$ & $\num[math-rm=\mathbf]{4.49e-1}\pm\num[math-rm=\mathbf]{4.27e-2}$ & $\num{3.78e-1}\pm\num{1.09e-2}$ & $\num{5.43e-1}\pm\num{2.57e-2}$ & $\num{-3.10e-2}\pm\num{1.91e-2}$ & $\num{6.11e-1}\pm\num{9.68e-3}$ & $\num{-2.65e-1}\pm\num{1.84e-2}$ & $\num{6.31e-1}\pm\num{2.64e-3}$ \\ 
\text{006\_S\_6689} &  $\num[math-rm=\mathbf]{8.27e-1}\pm\num[math-rm=\mathbf]{1.23e-2}$ & $\num[math-rm=\mathbf]{3.62e-1}\pm\num[math-rm=\mathbf]{8.80e-3}$ & $\num{8.12e-1}\pm\num{1.37e-2}$ & $\num{3.74e-1}\pm\num{2.38e-2}$ & $\num{3.59e-1}\pm\num{3.28e-2}$ & $\num{6.35e-1}\pm\num{1.01e-2}$ & $\num{4.24e-1}\pm\num{3.11e-2}$ & $\num{5.92e-1}\pm\num{9.34e-3}$ \\  \bottomrule
\end{tabular}
\end{adjustbox}
\label{tab:clinical_inversion}
\end{table*}

\begin{figure*}[!t]
\centering
\includegraphics[width=1\textwidth]{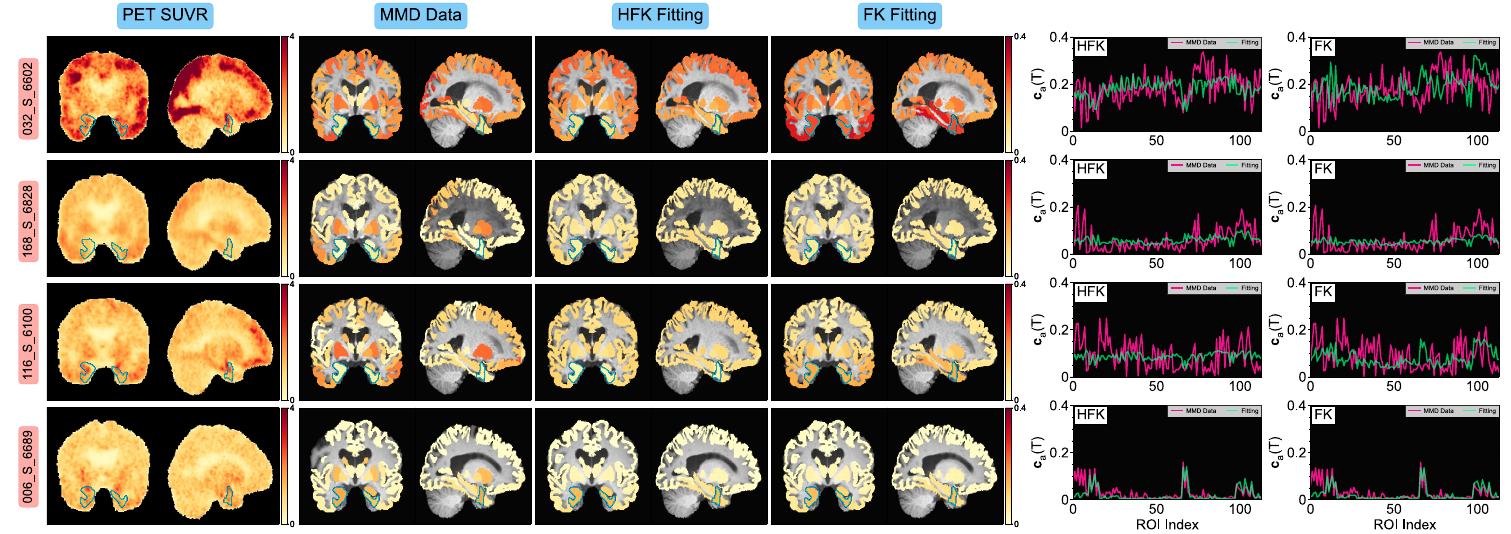}
\caption{Reconstructions of four clinical subjects with fixed IC. We assume the EC is the IC for all subjects here, and only try to estimate the scalar parameters. Each row stands for one clinical subject. From left to right in each panel, the figure shows SUVR image, processed regional tau abnormal MMD values (which is used as the inversion data), fitting from HFK model, fitting from FK model, 1D curve of fitting result from HFK model and FK model. The EC region is highlighted by solid contour line.}
\label{fig:clinical_inversion_sparsityEC}
\end{figure*}

\begin{figure*}[!t]
\centering
\includegraphics[width=1\textwidth]{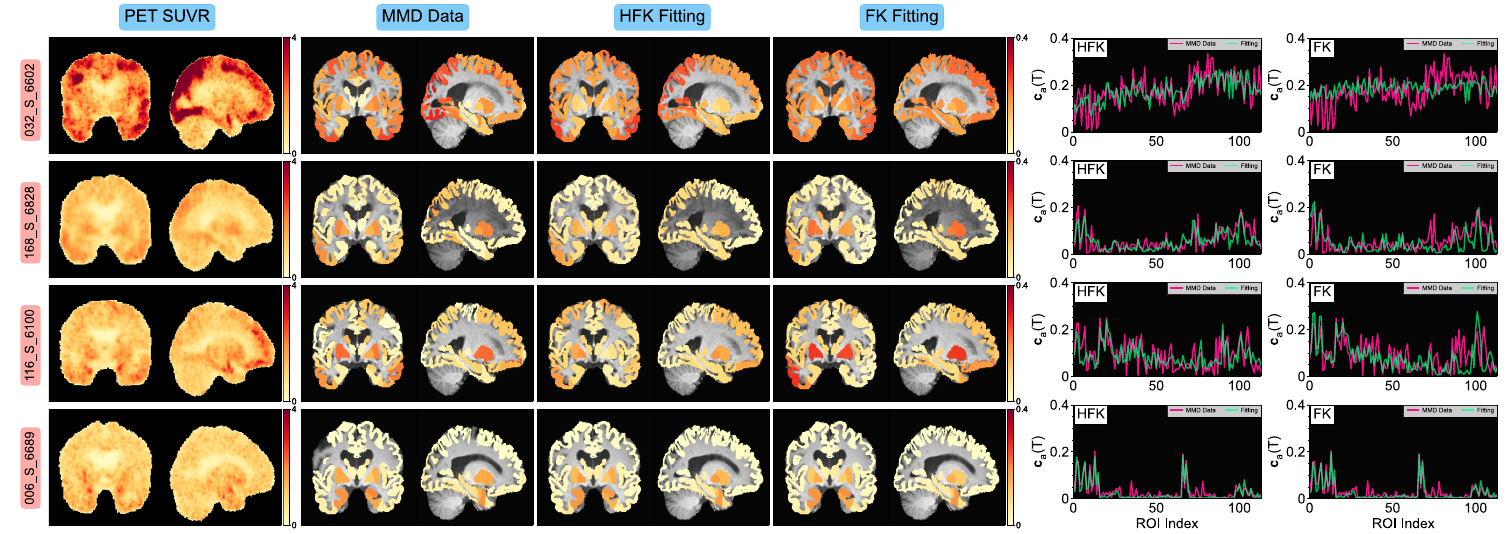}
\caption{Reconstructions of four clinical subjects. Instead of assuming EC as initial condition, we try to estimate the sparse IC from observation and assume the sparsity level of IC $s^{\text{max}}=5$. Each row stands for one clinical subject. From left to right in each panel, the figure shows PET image, processed regional tau abnormal MMD values (which is observation during the inversion), fitting from HFK model, fitting from FK model, 1D curve of fitting result from HFK model and FK model.}
\label{fig:clinical_inversion_sparsity5}
\end{figure*}

\noindent\textbf{(CQ1)} \textit{Evaluation of the MMD abnormality indicator: }
We demonstrate that regional MMD feature is a better indicator of tau abnormality than regional SUVR or tau positive probability in \cite{vogel2020spread}. To accomplish this, we take 76 cognitively normal (CN) subjects with age below 65 from the ADNI dataset and 45 subjects with Alzheimer's disease (AD). The AD population has ages $74.96\pm8.87$ and 19 are females, and 26 are males. CN subjects are younger with ages $61.76\pm3.71$ with 59 females and 17 males. For each subject, we extract regional SUVR, tau probability  \cite{vogel2020spread} and MMD scores. We explore three scenarios: we perform binary classification using a support vector machine (SVM) with a polynomial kernel, utilizing SUVR features. We repeat this classification using tau probability and MMD scores. The dataset is split into 60\% training and 40\% testing for reliable evaluation.

To select the most relevant features for classification, we rank the features based on their Pearson correlation score with the subject labels. We then classify the subjects using a subset of the most highly correlated features. During the feature selection process, we incrementally add one additional feature each time and employ five-fold cross-validation to compute the mean and standard deviation of the classification accuracy on the training dataset. The optimal number of features is determined by selecting the subset with the highest mean-to-standard deviation ratio.

This feature selection results in using  four features for MMD, seven features for mean SUVR and twenty features for tau probability. We conduct the classification task and report the best-performing model for all three features in \Cref{tab:classification}. The MMD approach achieved 97.9\% accuracy on the test dataset compared to 89.4\% accuracy using SUVR values and 91.5\% using tau probability.

\begin{figure}[!t]
\centering
\includegraphics[width=1\linewidth]{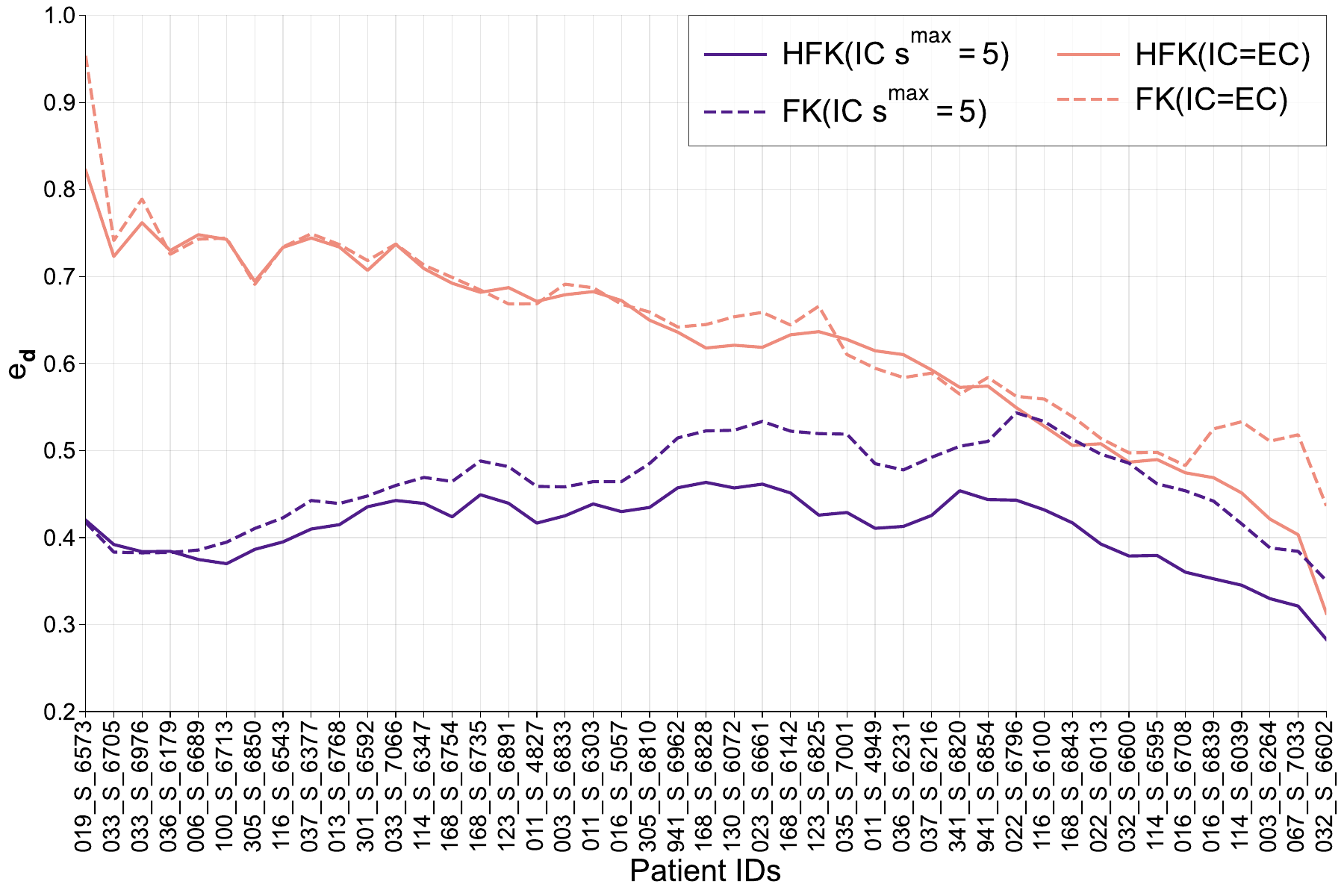}
\caption{Inversion for 45 AD subjects from ADNI dataset using the HFK and FK models in two scenarios: IC=EC and IC ($s^\max$=5). The FK method with IC=EC is a popular model that people use to reconstruct the tau-PET data. We compare the methods regarding $e_{\bold{d}}$. The lines are moving average of relative errors across adjacent two subjects. The subjects are sorted by their summation of MMD concentration. The proposed inversion algorithm (IC ($s^\max$=5)) method with HFK model overall performs the best.}
\label{fig:clinical_inversion_summary}
\end{figure}

\begin{figure}[!t]
\centering
\includegraphics[width=1\linewidth]{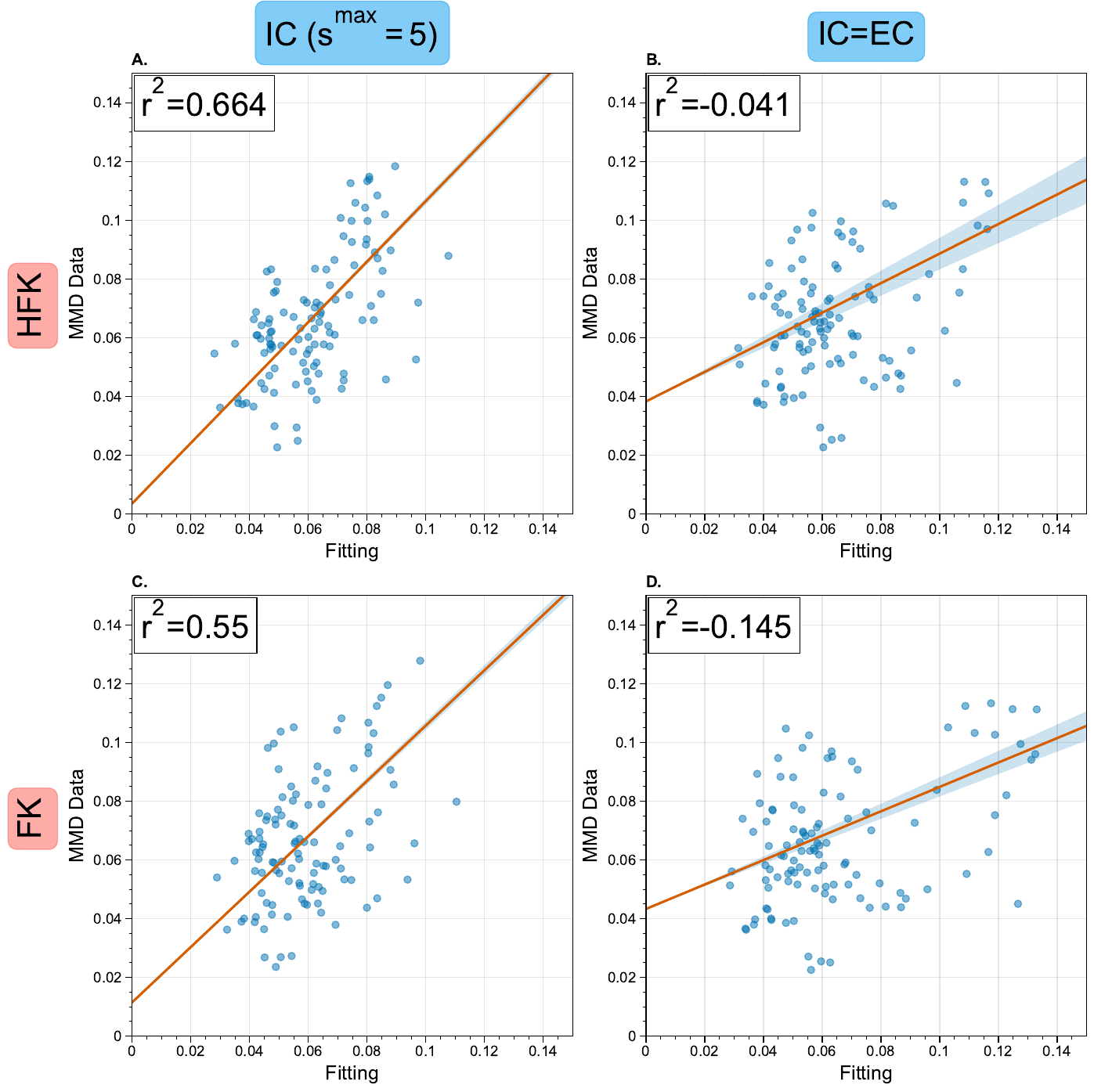}
\caption{$r^2$ score for AD population using HFK and FK models with IC ($s^\max$=5) and IC=EC inversion methods. We compute the $r^2$ between averaged observation regional MMD and averaged fitting result for AD subjects. Each dot represents average value at one parcel among all the population. We fit the dots by affine function and the blue shadow represents the three times standard deviation of fitted curve.}
\label{fig:clinical_r2}
\end{figure}

\noindent\textbf{(CQ2)} \textit{HFK and FK model comparison: }We conduct two experiments on four manually selected patients from the ADNI dataset, each with different tau distributions. In the first experiment, we run inversions with the HFK and FK models, assuming the initial condition (IC) is located at EC (IC=EC), and estimated only the parameters $\kappa$, $\rho$, and $\gamma$.
Second, we use our inversion algorithm for HFK and FK to estimate the parameters $\kappa$, $\rho$, $\gamma$ and $\bold{p}_0$. In the second experiment, as part of our inversion \Cref{subsection: Inverse}, we incorporate a sparsity hyper-parameter $s^{\text{max}}$. We select $s^{\text{max}}=5$ based on our tests, as the model's performance does not significantly change as we increase it. For the rest of the paper, we refer to the algorithm with IC inversion and $s^{\text{max}}=5$ as IC ($s^\max$=5), while the inversion with a fixed IC at EC is denoted as IC=EC.

In \Cref{fig:clinical_inversion_sparsityEC} and \Cref{fig:clinical_inversion_sparsity5}, we plot the first and second scenarios, respectively, for four subjects. We visualize the SUVR values for tau-PET scans, the parcellated tau MMD score (referred to as "MMD data"), and the inversion results (Fitting) obtained using the FK and HFK models. We summarize the fitting performance for each subject in \Cref{tab:clinical_inversion}, where we report the $\ell_2$ relative error ($e_{\mathbf{d}}$) and $r^2$ values for both experiments. We assume the tau-PET image follows the voxel-wise Gaussian distribution with mean and standard deviation extracted from \Cref{subsubsection: processing}. We take subject tau-PET samples from this distribution. We report both mean and standard deviation as for the sensitivity evaluation of inversion regarding samples for each subject. For the case with IC=EC, in comparison to the FK model, the HFK model achieves upto 15.4\% improvement in the $\ell_2$ norm relative error. In the case with IC ($s^\max$=5), HFK achieves a 27.3\% relative error improvement (measured in $\ell_2$ norm) over the HFK model with IC=EC. Both FK and HFK models inverting IC outperform the cases with IC=EC, as they provide a greater degree of freedom in IC selection. The HFK model in which we invert for IC exhibits the best overall performance.
\par\medskip

\begin{figure}[!t]
\centering
\includegraphics[width=1\linewidth]{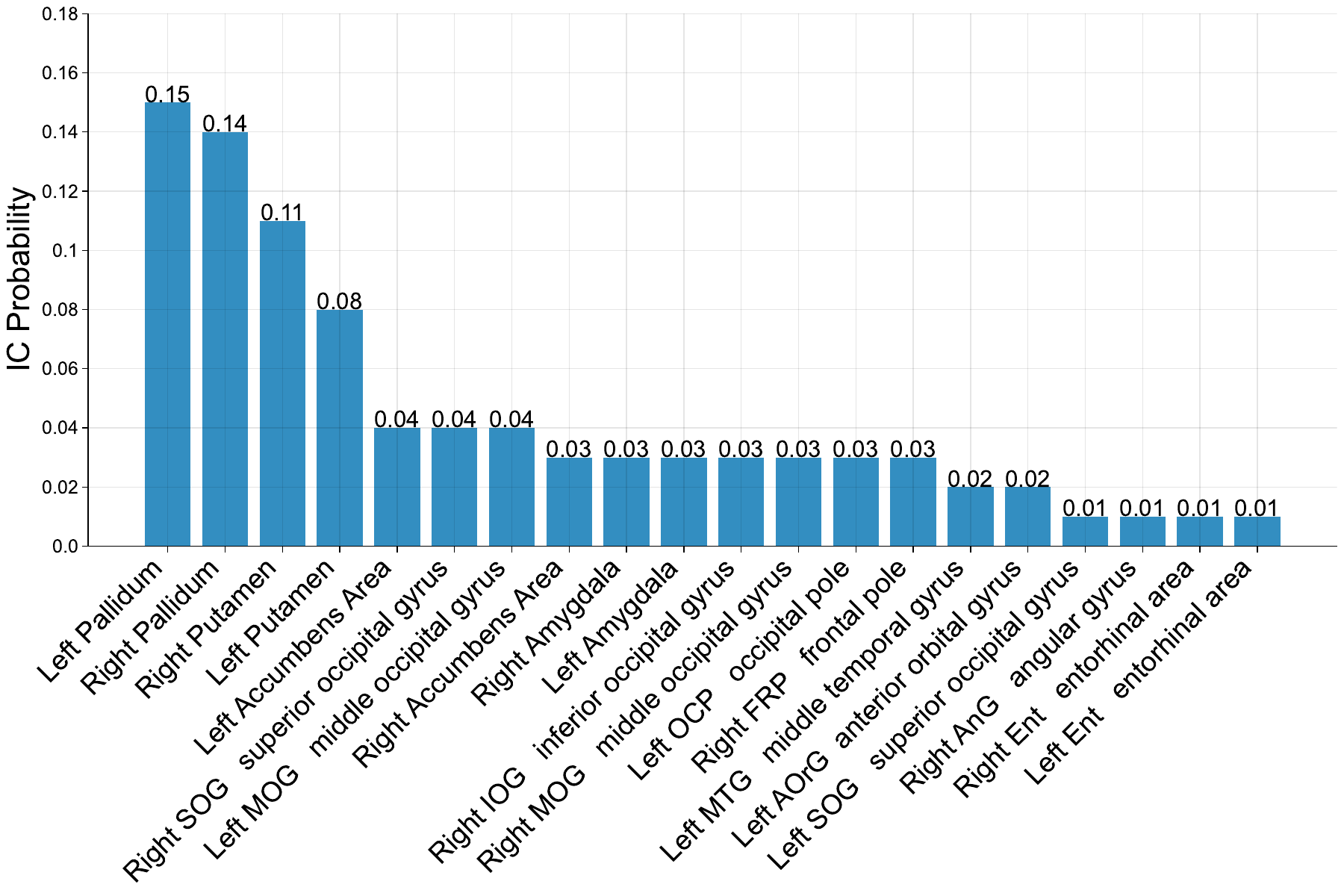}
\caption{Probability of regions to be IC of AD. We pick the 20 mostly selected IC through our inversion for clinical AD subjects. Under the presence of high variability in AD tau-PET data, EC is less likely to be chosen as the IC of disease propagation with probability 1\%. The most common choice is Pallidum with probability as high as 15\%.}
\label{fig:clinical_ic_histogram}
\end{figure}

\begin{figure}[!t]
\centering
\includegraphics[width=1\linewidth]{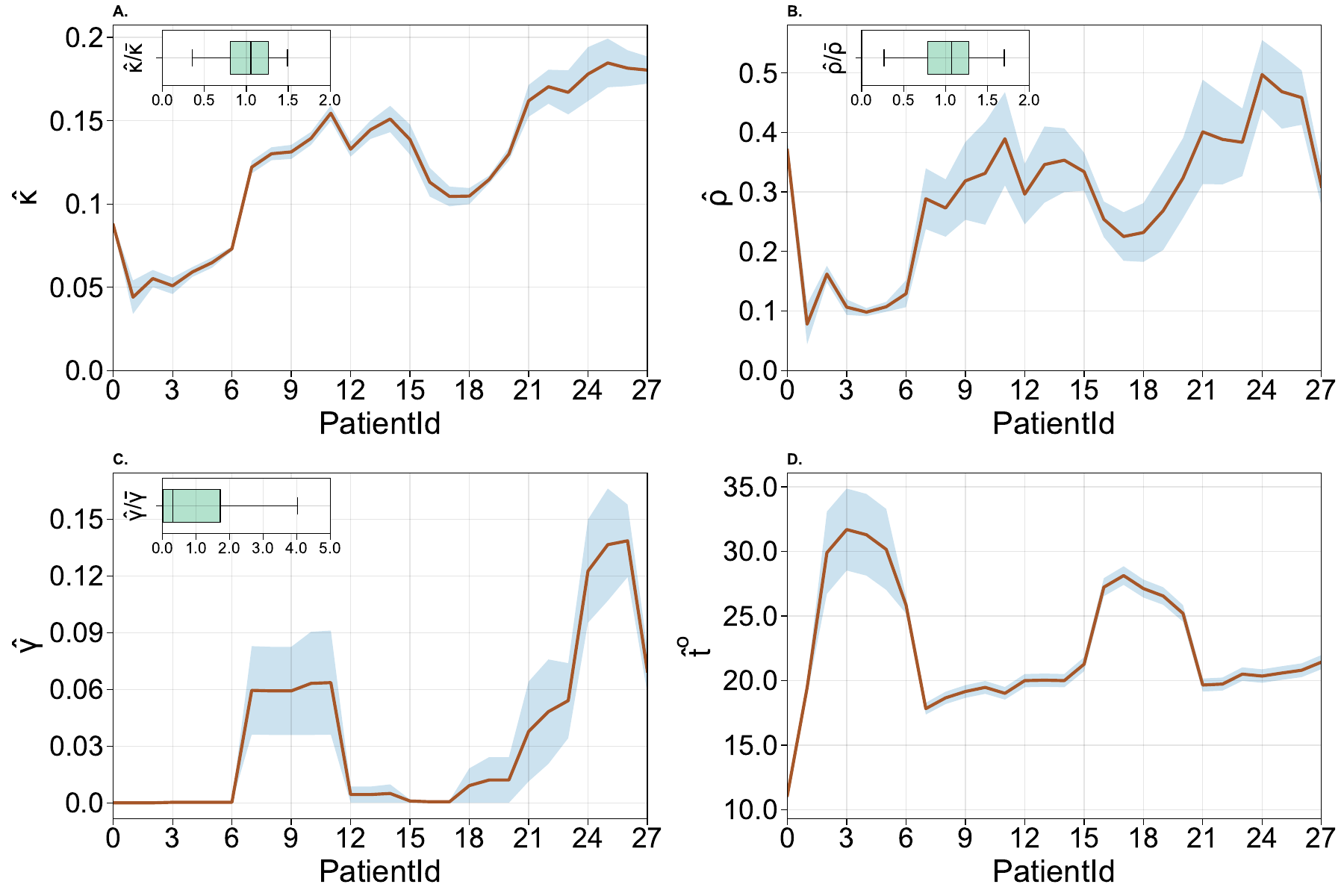}
\caption{Longitudinal analysis of clinical patients. The line plots with blue shaded regions represent the means ($\hat{\kappa}$, $\hat{\rho}$, $\hat{\gamma}$ and $\hat{\mathsf{t}}^o$) and standard deviations ($\sigma_{\kappa}$, $\sigma_{\rho}$, $\sigma_{\gamma}$ and $\sigma_{\mathsf{t}^o}$) of the inverted parameters. Low standard deviations of inverted parameters for the same patient demonstrate the consistency of the inversion system. The box plots within the subplots summarize the population distributions of $\frac{\hat{\kappa}}{\bar{\kappa}}$, $\frac{\hat{\rho}}{\bar{\rho}}$, and $\frac{\hat{\gamma}}{\bar{\gamma}}$, where $\bar{\kappa}$, $\bar{\rho}$ and $\bar{\gamma}$ represent population averages of inverted parameters. In the bottom right subplot, we present the estimated times of disease onset in years. Most subjects are estimated to have had disease onset between 10 and 35 years prior to their scans.}
\label{fig:clinical_longitudinal_plot}
\end{figure}

\noindent\textbf{(CQ3)} \textit{Evaluation on ADNI dataset:}
In this section, we evaluate the performance of the HFK and FK models on a cohort of 455 CN, 212 MCI and 45 AD subjects from ADNI dataset. Here we mainly present results of AD subjects us tau distribution. We assess each model using fixed  IC=EC and unknown IC, which we indicate as IC ($s^\max = 5$).

In \Cref{fig:clinical_inversion_summary}, we present the relative error $e_{\mathbf{d}}$ for each subject. The subjects are ordered increasingly based on the level of tau abnormality ($\|\mathbf{d}\|_1$). In nearly all AD subjects, the HFK model with IC ($s^\max$=5) demonstrates superior performance. The HFK model achieves relative fitting errors as low as 28\%. Notably, for subject 019\_S\_6573, IC ($s^\max$=5) achieves $\sim 40\%$ improvement w.r.t IC=EC for HFK model. \Cref{fig:clinical_r2} demonstrates the $r^2$ for two models and two inversion schemes. HFK model with IC inversion can explain 66.4\% of the observation data, while the popular FK (EC) fails in the AD data fitting (with $r^2=-0.145$). We further test our scheme on 455 CN subjects and achieve $r^2=0.8$, and on 212 MCI subjects with $r^2=0.729$, which validates the effectiveness of our method in different cognitive group. Furthermore, we test the case that uses tau positive probability \cite{vogel2020spread} as observation, and adapt our inversion scheme. Tau probability achieves $r^2=0.5$ for AD subjects using $s^{\text{max}}=10$ which shows the superior potential of MMD over tau probability.

Now that we invert for the IC location, a question arises: is there a pattern on the initiation locations? And is the hypotheses of abnormality initiation at EC valid? We report the frequency of IC locations chosen by our inversion scheme in \Cref{fig:clinical_ic_histogram}. The results indicates a preference for the Pallidum ($\sim15\%$) and Putamen ($\sim10\%$), while the Entorhinal Cortex (EC) are chosen only around 1\%.
\par\medskip

\noindent\textbf{(CQ4)} \textit{Sensitivity analysis of inversion scheme: } 
To assess the sensitivity of the inversion scheme, we conduct experiments using different templates, parcellation, and connectivity matrices. For each subject, their T1 image is registered with five different template T1 images, resulting in variations in parcel size and location. Additionally, the T1 images reveal differences in grey matter and ventricle sizes across the brain. These variations directly impact the MMD score, as it depends on samples drawn from different regions. The inversion sensitivity regarding the five brain templates on AD subjects is within 5\% regarding $e_{\bold{d}}$.

To analyze the effects of parcellation on the model, we divide large regions into finer-grained regions via K-means to achieve a more uniform distribution of parcel volumes. There are around 450 parcels in the new parcellation (varies between templates). Using the new parcellation, we calculate new MMD scores for all the templates we have. We run the inversion algorithm for the new parcellation. HFK (IC ($s^\max$=5)) model achieves an $r^2$ score of $0.584$ using the finer parcellation.

Our connectivity matrix is compared to the one derived from the widely used MRtrix3 software \cite{tournier2019mrtrix3}. The inversion results are all done using HFK model with IC ($s^\max$=5) inversion method. Our connectivity matrix achieves slightly better fitting relative error (40.97\%) than MRtrix3 (45.18\%). But the main conclusion is that the inversion algorithm is stable to using different tractography methods.
\par\medskip

\noindent\textbf{(CQ5)} \textit{Inversion using longitudinal data: } For patients with more than two tau scans, we select the two most recent scans, with their chronological ages denoted as $\mathsf{t}_1$ and $\mathsf{t}_2$ (where $\mathsf{t}_1 < \mathsf{t}_2$). We compute the MMD scores of these two tau scans, denoted as $\bold{d}_1$ and $\bold{d}_2$, respectively. Our aim is to examine whether the inverted parameters are stable when using two tau scans from the same patient. Besides, we want to determine whether we can infer the disease onset time using longitudinal data and our inversion algorithm. We then perform the following four steps:
\begin{enumerate}
\item We run the inversion algorithm using $\bold{d}_1$. Given the inverted model parameters $\kappa_1$, $\rho_1$ and $\gamma_1$ and IC, we run the forward model in \Cref{eq: diffusion-pde} to obtain $\bold{c}_{a, 1}(t)$ for $t\in[1, 2]$. We then compute $\tau_1 = \arg\min_{t\in[1, 2]} \frac{\left\|\bold{c}_{a, 1}(t) - \bold{d}_1\right\|_2}{\left\|\bold{d}_1\right\|_2}$.
\item We follow a similar procedure using $\bold{d}_2$. Given the inverted model parameters $\kappa_2$, $\rho_2$ and $\gamma_2$ and IC, we run the forward model in \Cref{eq: diffusion-pde} to obtain $\bold{c}_{a, 2}(t)$ for $t\in[0, 1]$. We compute $\tau_2 = \arg\min_{t\in[0, 1]} \frac{\left\|\bold{c}_{a, 2}(t) - \bold{d}_2\right\|_2}{\left\|\bold{d}_2\right\|_2}$.
\item We estimate the disease onset times as $\hat{\mathsf{t}}^{o}_{1} = \frac{\mathsf{t}_2 - \mathsf{t}_1}{\tau_1 - 1}$ and $\hat{\mathsf{t}}^{o}_{2} = \frac{\mathsf{t}_2 - \mathsf{t}_1}{1 - \tau_2}$. We then report the average and standard deviation of these two estimates: $\hat{\mathsf{t}}^{o} = \frac{\hat{\mathsf{t}}^o_{1} + \hat{\mathsf{t}}^o_{2}}{2}$, $\sigma_{{\mathsf{t}}^o} = \left|\frac{\hat{\mathsf{t}}^{o}_{2} - \hat{\mathsf{t}}^{o}_{1}}{2}\right|$. 
\item We normalize the model parameters by computing $\hat{\kappa}_1={\kappa_1}/{\hat{\mathsf{t}}^o_{1}}$, $\hat{\rho}_1={\rho_1}/{\hat{\mathsf{t}}^o_{1}}$, $\hat{\gamma}_1={\gamma_1}/{\hat{\mathsf{t}}^o_{1}}$, and similarly for the second set of parameters: $\hat{\kappa}_2={\kappa_2}/{\hat{\mathsf{t}}^o_{2}}$, $\hat{\rho}_2={\rho_2}/{\hat{\mathsf{t}}^o_{2}}$, $\hat{\gamma}_2={\gamma_2}/{\hat{\mathsf{t}}^o_{2}}$. We then report the averages and standard deviations of the normalized parameters: $\hat{\kappa} = \frac{\hat{\kappa}_1 + \hat{\kappa}_2}{2}$, $\sigma_{{\kappa}}=\left|\frac{\hat{\kappa}_1 - \hat{\kappa}_2}{2}\right|$, $\hat{\rho} = \frac{\hat{\rho}_1 + \hat{\rho}_2}{2}$, $\sigma_{\rho} = \left|\frac{\hat{\rho}_1 - \hat{\rho}_2}{2}\right|$, $\hat{\gamma} = \frac{\hat{\gamma}_1 + \hat{\gamma}_2}{2}$ and $\sigma_{\gamma} = \left|\frac{\hat{\gamma}_1 - \hat{\gamma}_2}{2}\right|$.
\end{enumerate}
In \Cref{fig:clinical_longitudinal_plot}, we present $\hat{\kappa}$, $\hat{\rho}$, $\hat{\gamma}$ and $\hat{\mathsf{t}}^o$ in line plots. The blue shading in each subplot represents the standard deviations $\sigma_{\kappa}$, $\sigma_{\rho}$, $\sigma_{\gamma}$ and $\sigma_{\mathsf{t}^o}$. Patients are ranked in ascending order based on their abnormality scores, defined as $\left\|\bold{d}_1\right\|_1$. The results indicate that the majority of subjects experience disease onset between 10 and 35 years prior. Moreover, the low standard deviations demonstrate the consistency of the estimated parameters for the same patient, highlighting the stability of the inversion algorithm when using clinical data. Let $P$ denote the number of patients, and let $\hat{\kappa}^i$, $\hat{\rho}^i$ and $\hat{\gamma}^i$ be the normalized parameters for the $i^{\text{th}}$ patient ($i=1,\dots, P$). We compute the population averages: $\bar{\kappa}={1}/{P}\sum_{i}\hat{\kappa}^i$, $\bar{\rho}={1}/{P}\sum_{i}\hat{\rho}^i$, $\bar{\gamma}={1}/{P}\sum_{i}\hat{\gamma}^i$. We show the distributions of ${\hat{\kappa}}/{\bar{\kappa}}$, ${\hat{\rho}}/{\bar{\rho}}$ and ${\hat{\gamma}}/{\bar{\gamma}}$ using box plots in \Cref{fig:clinical_longitudinal_plot}. These box plots represent the variability of the inverted parameters across different patients, illustrating the inherent heterogeneity among AD patients.
\par\medskip

%% file: conclusion.tex
\section{Conclusion}
We propose a novel inversion algorithm for a two-species HFK model exhibits improved qualitative consistency with tau-PET data compared to the existing FK model. Our model utilizes sparse initial conditions and three scalar parameters to represent migration, proliferation, and clearance. To estimate these parameters, we employ an inversion strategy that utilizes quasi-newton and compressive sampling method.

We validate our algorithm under synthetic scenario, and evaluate on ADNI dataset focusing on 45 AD patients. Based on several criteria, such as relative error and $r^2$, the HFK model with inverted initial conditions outperforms other methods overall using only a small number of parameters. To verify the stability and effectiveness of the model, we conduct sensitivity experiments with templates, parcellations, and connectivity matrices. Additionally, we analyze longitudinal data using our inversion algorithm. By incorporating patient-specific model parameters, we estimate the time of onset for AD. Our model suggests the existence of additional seeding locations apart from the entorhinal cortex (EC).

However, our current work has some limitations. We focus solely on tau lesion modeling and do not consider the influence of $A\beta$, which is recognized as an important factor in AD. Future research endeavors will integrate the effect of $A\beta$ into the tau modeling. From the perspective of the inversion, it will be interesting to add sparsity constraints to the connectivity matrix and make the connectivity as an additional optimization parameter.

%% file: acknowledgement_data_availability.tex
\section{Ackwonledgement}
This material is based upon work supported by NSF award OAC 22042261; 
by the U.S. Department of Energy, Office of Science, Office of Advanced Scientific Computing Research, Applied Mathematics program, Mathematical Multifaceted Integrated Capability Centers (MMICCS) program, under award number DE-SC0023171; 
and by the U.S. National Institute on Aging under award number R21AG074276-01. 
Any opinions, findings, and conclusions or recommendations expressed herein are those of the authors and do not necessarily reflect the views of the DOE, NIH, and NSF. 
Computing time on the Texas Advanced Computing Centers Stampede system was provided by an allocation from TACC and the NSF. 

\section{Data Availability}
The data used in this study were obtained from the Alzheimer's Disease Neuroimaging Initiative (ADNI) database\cite{petersen2010alzheimer}.